\documentclass[12pt]{article}
\usepackage{epsfig}

\newcommand\pubnumber{ILL-(TH)-01-1}
\newcommand\pubdate{\today}
\newcommand\hepnumber{hep-ph/0103033}

\def\csumb{Department of Physics \\
University of Illinois at Urbana-Champaign \\ 1110 W.~Green St., Urbana, IL
61801}
\def\Title#1{\begin{center} {\Large\bf #1 } \end{center}}
\def\Author#1{\begin{center}{ \sc #1} \end{center}}
\def\Address#1{\begin{center}{ \it #1} \end{center}}

\newcommand\pubblock{\rightline{\begin{tabular}{l} \pubnumber\\
         \pubdate\\ \hepnumber \end{tabular}}}
\newenvironment{Abstract}{\begin{quotation}  }{\end{quotation}}
\newenvironment{Presented}{\begin{quotation} \begin{center}
             Presented at the\end{center}
      \begin{center}\begin{large}}{\end{large}\end{center} \end{quotation}}
\def\Acknowledgments{\bigskip  \bigskip \begin{center}
          \large\bf Acknowledgments\end{center}}

\makeatletter
\def\section{\@startsection{section}{0}{\z@}{5.5ex plus .5ex minus
 1.5ex}{2.3ex plus .2ex}{\large\bf}}
\def\subsection{\@startsection{subsection}{1}{\z@}{3.5ex plus .5ex minus
 1.5ex}{1.3ex plus .2ex}{\normalsize\bf}}
\def\subsubsection{\@startsection{subsubsection}{2}{\z@}{-3.5ex plus
-1ex minus  -.2ex}{2.3ex plus .2ex}{\normalsize\sl}}

\renewcommand{\@makecaption}[2]{%
   \vskip 10pt
   \setbox\@tempboxa\hbox{\small #1: #2}
   \ifdim \wd\@tempboxa >\hsize     % IF longer than one line:
       \small #1: #2\par          %   THEN set as ordinary paragraph.
     \else                        %   ELSE  center.
       \hbox to\hsize{\hfil\box\@tempboxa\hfil}
   \fi}

 \def\citenum#1{{\def\@cite##1##2{##1}\cite{#1}}}
\newcount\@tempcntc
\def\@citex[#1]#2{\if@filesw\immediate\write\@auxout{\string\citation{#2}}\fi
  \@tempcnta\z@\@tempcntb\m@ne\def\@citea{}\@cite{\@for\@citeb:=#2\do
    {\@ifundefined
       {b@\@citeb}{\@citeo\@tempcntb\m@ne\@citea\def\@citea{,}{\bf ?}\@warning
       {Citation `\@citeb' on page \thepage \space undefined}}%
    {\setbox\z@\hbox{\global\@tempcntc0\csname b@\@citeb\endcsname\relax}%
     \ifnum\@tempcntc=\z@ \@citeo\@tempcntb\m@ne
       \@citea\def\@citea{,}\hbox{\csname b@\@citeb\endcsname}%
     \else
      \advance\@tempcntb\@ne
      \ifnum\@tempcntb=\@tempcntc
      \else\advance\@tempcntb\m@ne\@citeo
      \@tempcnta\@tempcntc\@tempcntb\@tempcntc\fi\fi}}\@citeo}{#1}}
\def\@citeo{\ifnum\@tempcnta>\@tempcntb\else\@citea\def\@citea{,}%
  \ifnum\@tempcnta=\@tempcntb\the\@tempcnta\else
  {\advance\@tempcnta\@ne\ifnum\@tempcnta=\@tempcntb \else\def\@citea{--}\fi
    \advance\@tempcnta\m@ne\the\@tempcnta\@citea\the\@tempcntb}\fi\fi}
\makeatother

\def\beq{\begin{equation}}
\def\eeq#1{\label{#1}\end{equation}}
\def\eeqn{\end{equation}}

\newenvironment{Eqnarray}%
   {\arraycolsep 0.14em\begin{eqnarray}}{\end{eqnarray}}
\def\beqa{\begin{Eqnarray}}
\def\eeqa#1{\label{#1}\end{Eqnarray}}
\def\eeqan{\end{Eqnarray}}

\let\bar=\overbar

\def\Dslash{\not{\hbox{\kern-4pt $D$}}}
\def\dslash{\not{\hbox{\kern-2pt $\del$}}}

\def\msb{{\bar{\ssstyle M \kern -1pt S}}}

\def\lsim{\mathrel{\raise.3ex\hbox{$<$\kern-.75em\lower1ex\hbox{$\sim$}}}}
\def\gsim{\mathrel{\raise.3ex\hbox{$>$\kern-.75em\lower1ex\hbox{$\sim$}}}}

\begin{document}
\begin{titlepage}
\pubblock

\vfill
\def\thefootnote{\fnsymbol{footnote}}
\Title{Precision Top-Quark Physics}
\vfill \Author{Scott Willenbrock}
\Address{\csumb} \vfill
\begin{Abstract}
I consider the measurement of the top-quark mass, the CKM matrix element
$V_{tb}$, and the top-quark Yukawa coupling to the Higgs boson at the
Tevatron, the LHC, and a Linear Collider.  The theoretical motivations for
these measurements, as well as the experimental possibilities, are discussed.
\end{Abstract} \vfill
\begin{Presented}
5th International Symposium on Radiative Corrections \\
(RADCOR--2000) \\[4pt]
Carmel CA, USA, 11--15 September, 2000
\end{Presented}
\vfill
\end{titlepage}
\def\thefootnote{\arabic{footnote}}
\setcounter{footnote}{0}

\section{Introduction}

The top quark was discovered in 1995 by the CDF~\cite{Abe:1995hr} and
D0~\cite{Abachi:1995iq} experiments at Fermilab, during Run I of the Tevatron
$p\bar p$ collider ($\sqrt S = 1.8$ TeV, $\int {\cal L} dt \approx 100$
pb$^{-1}$). In Table \ref{topresults} I briefly summarize the top-quark
measurements made in Run I and compare them with the expectations from the
standard model.  The standard model does not predict the top-quark mass, but
it can be inferred indirectly from precision electroweak experiments
\cite{Groom:2000in}, and this indirect mass is in good agreement with the
measured mass.  The strong interaction of the top quark is probed by measuring
the top-quark cross section, which proceeds via the strong processes $q\bar q,
gg \to t\bar t$; this cross section is in good agreement with
next-to-leading-order QCD.  The weak interaction of the top quark is probed in
a variety of ways.  In the three-generation standard model, top decays almost
exclusively to bottom, as confirmed by experiment. The branching ratios of top
to longitudinal (zero-helicity) $W$ bosons and to right-handed
(positive-helicity) $W$ bosons are predicted to be approximately $0.7$ and
zero, respectively, in agreement with experiment. The weak interaction of the
top quark is also probed indirectly by precision electroweak experiments and
$b$-quark physics, as illustrated in Figs.~1 and 2. All of these experiments
are consistent with the three-generation standard model. Thus, although the
properties of the top quark have thus far been measured only crudely, there is
no evidence for physics beyond the standard model in top-quark physics.

%%%%%%%%%%%%%%%%%%%%%%%%%%%%%%%%%%%%%%%%%%%%%%%%%%%%%%%%%%%%%%%%%%%%%%%%%
%%
%%   use this format to include a LaTeX table  into your paper
%%
\def\tableline{\noalign{%\vskip-.5pt
\hrule height.7pt depth0pt\vskip3pt}}

\begin{table}[h]
\caption{Comparison of theory and experiment for top-quark physics from Run I
of the Fermilab Tevatron.  [$W_{0,+}$ denote a longitudinal (zero-helicity)
and right-handed (positive-helicity) $W$ boson.]  For discussion and
references, see
Refs.~\cite{Willenbrock:2000vv,Simmons:2000hr}.}\label{topresults}
\begin{center}
\setlength{\tabcolsep}{9pt}
\renewcommand{\arraystretch}{1.2}
\begin{tabular}{lll}
\tableline & Experiment & Theory \\ \tableline
$m_t$ & $174.3 \pm 5.1$ GeV & $168.2^{+9.6}_{-7.4}$ GeV \\
$\sigma(t\bar t)$ & $6.2 \pm 1.7$ pb & $4.75 \pm 0.5$ pb \\
$BR(t\to Wb)/BR(t\to Wq)$ & $0.94^{+0.31}_{-0.24}$ & $\approx 1$ \\
$BR(t\to W_0b)$ & $0.91 \pm 0.39$ & $\approx 0.7$ \\
$BR(t\to W_+b)$ & $0.11 \pm 0.15$ & $\approx 0$ \\
\hline
\end{tabular}
\end{center}
\end{table}

%%%%%%%%%%%%%%%%%%%%%%%%%%%%%%%%%%%%%%%%%%%%%%%%%%%%%%%%%%%%%%%%%%%%%%%%%

\begin{figure}[t]
\begin{center}
\epsfxsize= 5.5in \leavevmode \epsfbox{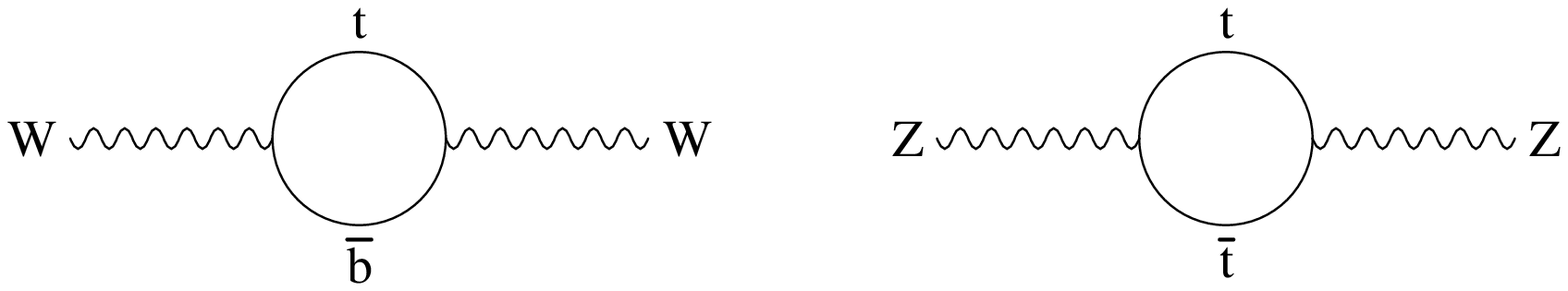}
\end{center}
\caption{The weak interaction of the top quark is probed indirectly by the
vector-boson self energies. }\label{selfenergy}
\end{figure}

\begin{figure}[t]
\begin{center}
\epsfxsize= 5.0in \leavevmode \epsfbox{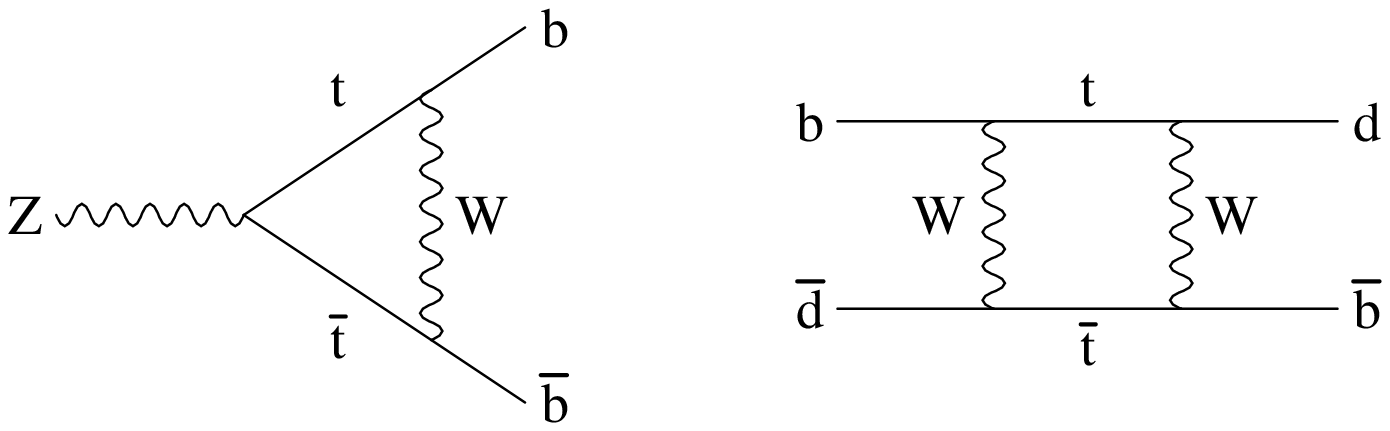}
\end{center}
\caption{The weak interaction of the top quark is also probed indirectly by
$b$ physics.}\label{virtualtop}
\end{figure}

Let us assume that the top quark is indeed a standard quark.  What parameters
of the top quark do we want to measure?  There are only a few standard-model
parameters associated with the top quark; its mass ($m_t$), its
Cabibbo-Kobayashi-Maskawa matrix elements ($V_{tb}$, $V_{ts}$, $V_{td}$), and
its Yukawa coupling to the Higgs field ($y_t$).  This last parameter is not
truly independent, as it is related to the top-quark mass (at leading order)
via $y_t = \sqrt 2 m_t/v$, where $v \approx 246$ GeV is the vacuum-expectation
value of the Higgs field.  However, this parameter is especially interesting,
as it is related to the electroweak-symmetry-breaking sector, which has yet to
be directly probed experimentally.  A measurement of the top-quark Yukawa
coupling therefore probes the mechanism that generates the top-quark mass.

In this talk I discuss the measurement of the parameters $m_t$, $V_{tb}$, and
$y_t$.  I first ask how accurately we desire these parameters.  I then ask how
accurately we can measure them with present and future colliders.  I consider
the upgraded Fermilab Tevatron ($\sqrt S = 2$ TeV), with an integrated
luminosity of 2 fb$^{-1}$ (Run IIa) and 15 fb$^{-1}$ (Run IIb), the CERN Large
Hadron Collider (LHC, $\sqrt S = 14$ TeV $pp$ collider), and the Linear
Collider, an $e^+e^-$ collider running at the $t\bar t$ threshold [as well as
at the $W^+W^-$ threshold and at the $Z$ mass (Giga $Z$)].

The standard-model parameters of the top quark are interesting in their own
right. Furthermore, any discrepancies between theory and experiment would
indicate new physics. Thus top-quark physics could serve to further solidify
the standard model, or to indicate physics beyond the standard model.

\begin{figure}[htb]
%\vspace{3 cm}
\begin{center}
% GNUPLOT: LaTeX picture
\setlength{\unitlength}{0.240900pt}
\ifx\plotpoint\undefined\newsavebox{\plotpoint}\fi
\sbox{\plotpoint}{\rule[-0.200pt]{0.400pt}{0.400pt}}%
\begin{picture}(974,809)(0,0)
\font\gnuplot=cmr10 at 10pt
\gnuplot
\sbox{\plotpoint}{\rule[-0.200pt]{0.400pt}{0.400pt}}%
\put(221.0,103.0){\rule[-0.200pt]{4.818pt}{0.400pt}}
\put(199,103){\makebox(0,0)[r]{0.001}}
\put(890.0,103.0){\rule[-0.200pt]{4.818pt}{0.400pt}}
\put(221.0,219.0){\rule[-0.200pt]{4.818pt}{0.400pt}}
\put(199,219){\makebox(0,0)[r]{0.01}}
\put(890.0,219.0){\rule[-0.200pt]{4.818pt}{0.400pt}}
\put(221.0,335.0){\rule[-0.200pt]{4.818pt}{0.400pt}}
\put(199,335){\makebox(0,0)[r]{0.1}}
\put(890.0,335.0){\rule[-0.200pt]{4.818pt}{0.400pt}}
\put(221.0,451.0){\rule[-0.200pt]{4.818pt}{0.400pt}}
\put(199,451){\makebox(0,0)[r]{1}}
\put(890.0,451.0){\rule[-0.200pt]{4.818pt}{0.400pt}}
\put(221.0,568.0){\rule[-0.200pt]{4.818pt}{0.400pt}}
\put(199,568){\makebox(0,0)[r]{10}}
\put(890.0,568.0){\rule[-0.200pt]{4.818pt}{0.400pt}}
\put(221.0,684.0){\rule[-0.200pt]{4.818pt}{0.400pt}}
\put(199,684){\makebox(0,0)[r]{100}}
\put(890.0,684.0){\rule[-0.200pt]{4.818pt}{0.400pt}}
\put(221.0,68.0){\rule[-0.200pt]{165.980pt}{0.400pt}}
\put(910.0,68.0){\rule[-0.200pt]{0.400pt}{167.907pt}}
\put(221.0,765.0){\rule[-0.200pt]{165.980pt}{0.400pt}}
\put(45,416){\makebox(0,0){$m_{q}$ (GeV)}}
\put(359,153){\makebox(0,0)[l]{$u$}}
\put(457,187){\makebox(0,0)[l]{$d$}}
\put(556,338){\makebox(0,0)[l]{$s$}}
\put(654,465){\makebox(0,0)[l]{$c$}}
\put(753,524){\makebox(0,0)[l]{$b$}}
\put(851,709){\makebox(0,0)[l]{$t$}}
\put(221.0,68.0){\rule[-0.200pt]{0.400pt}{167.907pt}}
\put(285,123){\usebox{\plotpoint}}
\put(285.0,123.0){\rule[-0.200pt]{16.622pt}{0.400pt}}
\put(354.0,123.0){\rule[-0.200pt]{0.400pt}{14.695pt}}
\put(285.0,184.0){\rule[-0.200pt]{16.622pt}{0.400pt}}
\put(285.0,123.0){\rule[-0.200pt]{0.400pt}{14.695pt}}
\put(383,158){\usebox{\plotpoint}}
\put(383.0,158.0){\rule[-0.200pt]{16.622pt}{0.400pt}}
\put(452.0,158.0){\rule[-0.200pt]{0.400pt}{13.490pt}}
\put(383.0,214.0){\rule[-0.200pt]{16.622pt}{0.400pt}}
\put(383.0,158.0){\rule[-0.200pt]{0.400pt}{13.490pt}}
\put(481,310){\usebox{\plotpoint}}
\put(481.0,310.0){\rule[-0.200pt]{16.863pt}{0.400pt}}
\put(551.0,310.0){\rule[-0.200pt]{0.400pt}{12.527pt}}
\put(482.0,362.0){\rule[-0.200pt]{16.622pt}{0.400pt}}
\put(482.0,310.0){\rule[-0.200pt]{0.400pt}{12.527pt}}
\put(580,456){\usebox{\plotpoint}}
\put(580.0,456.0){\rule[-0.200pt]{16.622pt}{0.400pt}}
\put(649.0,456.0){\rule[-0.200pt]{0.400pt}{2.891pt}}
\put(580.0,468.0){\rule[-0.200pt]{16.622pt}{0.400pt}}
\put(580.0,456.0){\rule[-0.200pt]{0.400pt}{2.891pt}}
\put(679,523){\usebox{\plotpoint}}
\put(679.0,523.0){\rule[-0.200pt]{16.622pt}{0.400pt}}
\put(748.0,523.0){\rule[-0.200pt]{0.400pt}{0.723pt}}
\put(679.0,526.0){\rule[-0.200pt]{16.622pt}{0.400pt}}
\put(679.0,523.0){\rule[-0.200pt]{0.400pt}{0.723pt}}
\put(777,708){\usebox{\plotpoint}}
\put(777.0,708.0){\rule[-0.200pt]{16.622pt}{0.400pt}}
\put(846.0,708.0){\rule[-0.200pt]{0.400pt}{0.723pt}}
\put(777.0,711.0){\rule[-0.200pt]{16.622pt}{0.400pt}}
\put(777.0,708.0){\rule[-0.200pt]{0.400pt}{0.723pt}}
\end{picture}
%\epsfxsize=2.5in
%\leavevmode
%\epsfbox[100 60 300 250]{massspectrum.eps}
\end{center}
\caption{The quark mass spectrum.  The bands indicate the running
$\overline{\rm MS}$ mass, evaluated at the quark mass (for $c,b,t$) or at 2
GeV (for $u,d,s$), and the associated uncertainty.} \label{massspectrum}
\end{figure}

\section{Top-quark mass}

The top-quark mass has been measured by the CDF and D0 collaborations to be
\cite{Demortier:1999vv}\footnote{This is the top-quark pole mass, which is
defined to order $\Lambda_{QCD} \approx 200$ MeV \cite{Smith:1997xz}. The
corresponding $\overline{\rm MS}$ mass is $m_t^{\overline {\rm
MS}}(m_t^{\overline {\rm MS}}) = 165.2 \pm 5.1$ GeV \cite{Gray:1990yh}.}
\begin{equation}
m_t = 174.3 \pm 5.1\;{\rm GeV}\;({\rm CDF} + {\rm D0})\;.
\end{equation}
To put this into context, I plot all the quark masses in
Fig.~\ref{massspectrum}, on a logarithmic scale. The width of each band is
proportional to the fractional uncertainty in the quark mass. We see that, at
present, the top-quark mass is the best-known quark mass, with the $b$-quark
mass a close second ($m_b^{\overline {\rm MS}}(m_b)=4.25 \pm 0.15$ GeV)
\cite{Groom:2000in}.

\begin{figure}[t]
\begin{center}
\epsfxsize= 3.0in \leavevmode \epsfbox{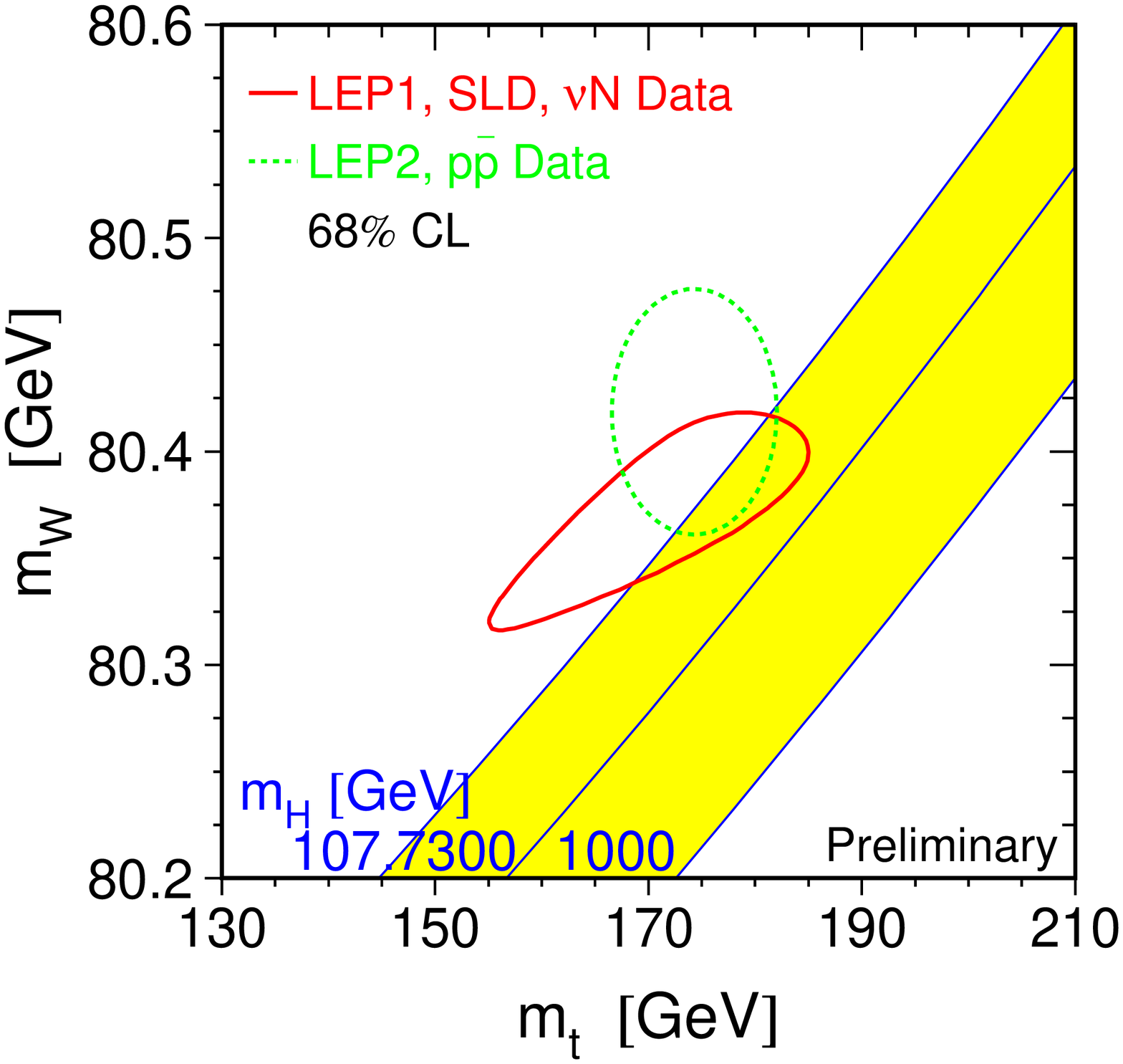}
\end{center}
\caption[fake]{$W$ mass {\it vs.}~top-quark mass, with lines of constant Higgs
mass. The solid ellipse is the $1\sigma$ ($68\%$ CL) contour from precision
electroweak experiments.  The dashed ellipse is the $1\sigma$ ($68\%$ CL)
contour from direct measurements.  Only the shaded region is allowed in the
standard electroweak model.  Figure from LEP Electroweak Working Group,
http://www.cern.ch/LEPEWWG/.}
\label{mwmt}
\end{figure}

An important question for the future is what precision we desire for the
top-quark mass.  There are at least two avenues along which to address this
question.  One is in the context of precision electroweak data.
Fig.~\ref{mwmt} summarizes the world's precision electroweak data on a plot of
$M_W$ {\it vs.}~$m_t$. The solid ellipse is the $1\sigma$ contour.  If the
standard electroweak model is correct, the predicted top-quark mass from
precision electroweak data is $m_t=168.2^{+9.6}_{-7.4}$ GeV
\cite{Groom:2000in}. We conclude that the present uncertainty of $5$ GeV in
the top-quark mass is sufficient for the purpose of precision electroweak
physics at this time.

There is one electroweak measurement, $M_W$, whose precision will increase
significantly. An uncertainty of $20$ MeV is a realistic goal for Run IIb at
the Tevatron and the LHC \cite{Amidei:1996dt,ATLAS,Beneke:2000hk}. Let us take
this uncertainty and project it onto a line of constant Higgs mass in
Fig.~\ref{mwmt}. This is appropriate, because once a Higgs boson is
discovered, even a crude knowledge of its mass will define a narrow line in
Fig.~3, since precision electroweak measurements are sensitive only to the
logarithm of the Higgs mass. An uncertainty in $M_W$ of $20$ MeV projected
onto a line of constant Higgs mass corresponds to an uncertainty of $3$ GeV in
the top-quark mass.  Thus we desire a measurement of $m_t$ to $3$ GeV in order
to make maximal use of the precision measurement of $M_W$ at the Tevatron and
the LHC.

Looking further ahead, a high-luminosity Linear Collider running at the $WW$
threshold could measure the $W$ mass with an accuracy of 6 MeV
\cite{Zerwas:2000gt}. This would require a measurement of $m_t$ to 1 GeV. The
same machine running at the $Z$ mass (Giga $Z$) could make a measurement of
$\sin^2\theta_W$ with an accuracy of $1\times 10^{-5}$ \cite{Zerwas:2000gt}.
This would also require a measurement of $m_t$ with an uncertainty of order 1
GeV \cite{Erler:2000jg}.

Another avenue along which to address the desired accuracy of the top-quark
mass is to recall that the top-quark mass is a fundamental parameter of the
standard model.  Actually, the fundamental parameter is the Yukawa coupling of
the top quark to the Higgs field, given at leading order by
\begin{equation}
y_t = \sqrt 2 \frac{m_t}{v} \approx 1\ \label{yukawa}
\end{equation}
where $v \approx 246$ GeV is the vacuum-expectation value of the Higgs field.
The fact that this coupling is of order unity suggests that it may be a truly
fundamental parameter.  We hope someday to have a theory that relates the
top-quark Yukawa coupling to that of its weak-interaction partner, the $b$
quark.\footnote{A particularly compelling model that relates the $b$ and $t$
masses is SO(10) grand unification \cite{Georgi:1975my,Fritzsch:1975nn}. This
model may be able to account for the masses of all the third-generation
fermions, including the tau neutrino, whose mass is given by the ``see-saw''
mechanism \cite{GRS} as $m_{\nu_\tau} \approx m_t^2/M_{GUT} \approx 10^{-2}$
eV \cite{Wilczek:1999cp}.} The $b$-quark mass is currently known with an
accuracy of $3.5\%$ \cite{Groom:2000in}. Since the uncertainty is entirely
theoretical, it is likely that it will be reduced in the future.  If we assume
that future work reduces the uncertainty to $1\%$, the corresponding
uncertainty in the top-quark mass would be $2$ GeV.

We conclude that both precision electroweak experiments and $m_t$ as a
fundamental parameter lead us to the desire to measure the top-quark mass with
an accuracy of 1--3 GeV.  This is well matched with future expectations.  An
uncertainty of 3 GeV per experiment is anticipated in Run IIa
\cite{Blair:1996kx,Abachi:1996hs}, and $2$ GeV per experiment in Run IIb
\cite{Amidei:1996dt}. The LHC could potentially reduce the uncertainty to 1
GeV, although that has not been established \cite{ATLAS}.

Recall that the need to reduce the uncertainty in the top-quark mass to 1 GeV
is driven by the precision measurement of $M_W$ and $\sin^2\theta_W$ at the
Linear Collider. Such a machine, operating at the $t\bar t$ threshold, could
make a much more accurate determination of the top-quark mass.  It is
interesting to ask if there is any motivation to go beyond 1 GeV in the
accuracy of the measurement of $m_t$.

\begin{figure}[t]
\begin{center}
\epsfxsize= 2.5in \leavevmode \epsfbox{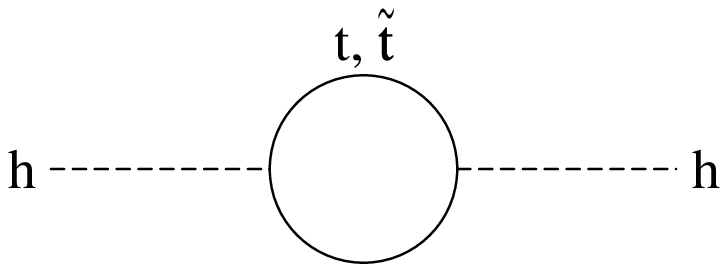}
\end{center}
\caption{Corrections to the mass of the lightest supersymmetric Higgs boson
from loops of top quarks and top squarks.}
\label{higgsselfenergy}
\end{figure}

No such motivation appears to exist solely within the context of the standard
model, but it is plausible that physics beyond the standard model could lead
us to desire $m_t$ with an accuracy much less than 1 GeV.  I offer two
examples.  Imagine that nature is supersymmetric, and the Higgs sector
consists of two Higgs doublets, as in the minimal supersymmetric standard
model. There is an upper bound on the mass of the lightest Higgs scalar, and
this bound is saturated in the limit that the pseudoscalar Higgs mass and the
ratio of vacuum-expectation values, $\tan\beta$, are large.  The mass of the
lightest Higgs scalar is predicted to be
\cite{Haber:1991aw,Okada:1991vk,Ellis:1991nz}
\begin{equation}
m_h^2 = M_Z^2+\frac{3G_F}{\pi^2\sqrt 2}m_t^4\ln\frac{M_S^2}{m_t^2}
\label{mhsusy}
\end{equation}
where $M_S^2$ is the average of the two top-squark squared masses and I have
assumed no top-squark mixing, for simplicity. The second term is from loops of
top quarks and top squarks, as shown in Fig.~\ref{higgsselfenergy}, and since
it depends on the top-quark mass to the fourth power, an uncertainty in the
top-quark mass implies an uncertainty in the predicted Higgs mass.  The Higgs
mass will be measured with an accuracy of about $0.1\%$ at the LHC
\cite{ATLAS}; this requires a measurement of $m_t$ to about 100 MeV (where I
have taken $M_S \approx 1$ TeV).  However, there is an uncertainty in the
predicted Higgs mass due to the unknown three-loop contributions to
Eq.~(\ref{mhsusy}), which has been estimated to be about 3 GeV
\cite{Heinemeyer:1999np,Heinemeyer:2001pq}.  This corresponds to an uncertainty
in the top-quark mass of about 2 GeV.  Thus the motivation to go beyond 1 GeV
accuracy hinges on the knowledge of higher-order terms in Eq.~(\ref{mhsusy}).

A second motivation for going beyond 1 GeV in the accuracy of the top-quark
mass measurement is a more model-independent one.  The generation of mass is
related to the breaking of the electroweak symmetry.  The electroweak
interaction has been measured with an accuracy of about $0.07\%$ (the accuracy
in our present knowledge of $\sin^2\theta_W$).  If the mechanism that breaks
the weak interaction is related to the weak interaction itself, then a
measurement of $m_t$ to $0.07\%$, {\it i.e.}, 100 MeV, may be warranted.

\begin{figure}[htb]
\begin{center}
\epsfxsize= 4.0in \leavevmode \epsfbox{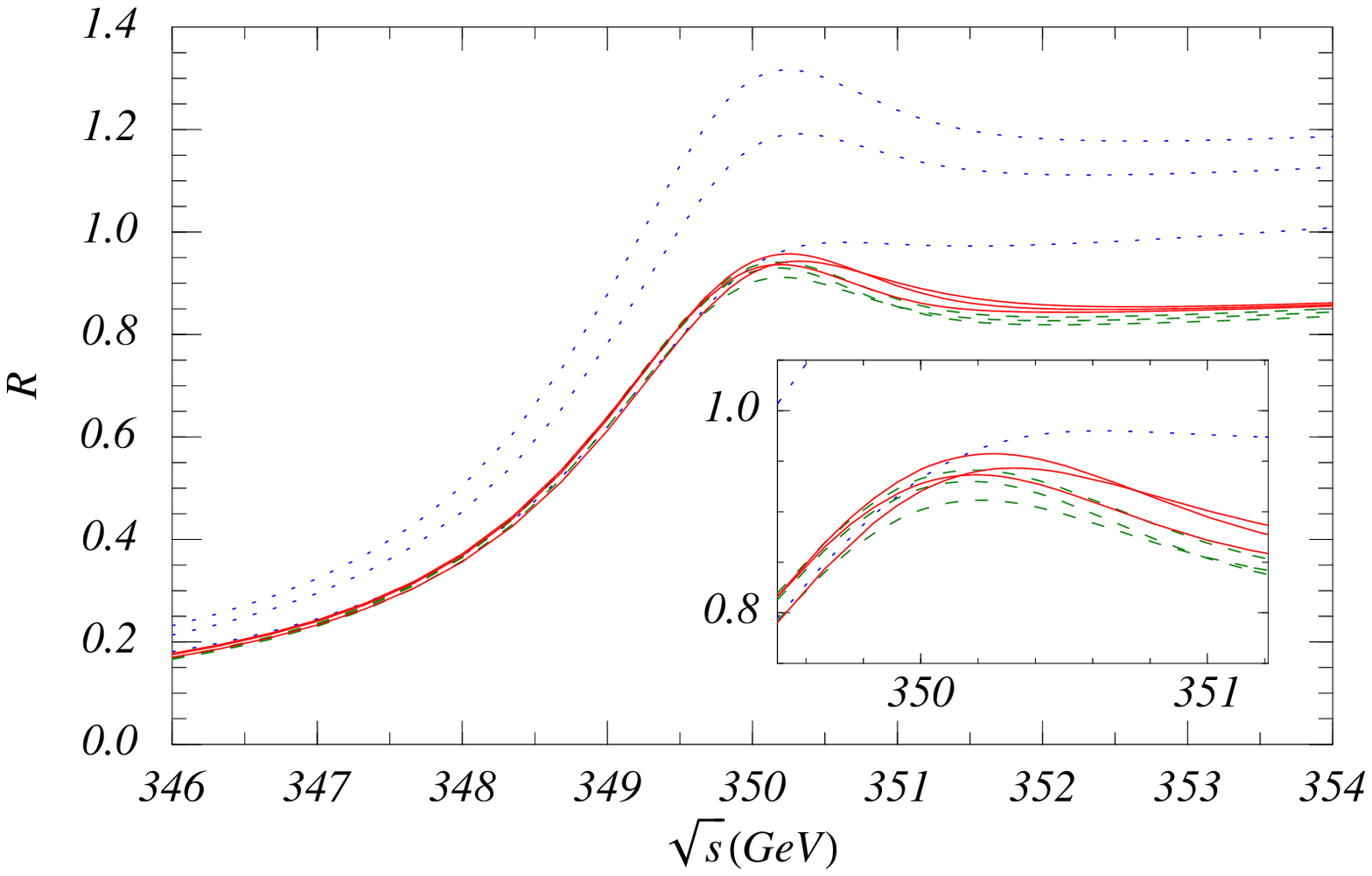}
\end{center}
\caption{Renormalization-group-improved NNLO calculation of the $t\bar t$
threshold at a Linear Collider.  The curves are leading log (dotted),
next-to-leading log (dashed), and next-to-next-to-leading log (solid), for
three different renormalization scales.  From
Ref.~\cite{Hoang:2000ib}.}\label{thresh}
\end{figure}

Both of these arguments, although speculative, lead to a goal of about 100 MeV
for the accuracy of the top-quark mass measurement.  Such an accuracy may be
within the reach of a Linear Collider operating at the $t\bar t$ threshold.
Recent next-to-next-to-leading-order (NNLO) calculations of the $t\bar t$
threshold in a nonrelativistic expansion yield a line shape with sufficient
accuracy to extract the mass within 100 MeV \cite{Hoang:2000yr}.  It is
essential to use a short-distance ``threshold mass'' in such calculations
\cite{Beneke:1998rk,Hoang:1999nz}. The threshold mass has recently been related
to the more commonly-used $\overline{\rm MS}$ mass to ${\cal O}(\alpha_s^3)$
\cite{Chetyrkin:2000qi,Melnikov:2000qh}, so the theoretical work required for
a NNLO extraction of the top-quark mass from the $t\bar t$ threshold is
complete. However, at the time of this symposium, there remained a mystery in
the normalization of the line shape. Work performed after this symposium has
resolved that mystery via renormalization-group improvement, as shown in
Fig.~\ref{thresh}, so the normalization is now also known with good accuracy
\cite{Hoang:2000ib}.

\section{$V_{tb}$}

It is remarkable that, although it has not yet been directly measured, $V_{tb}$
is the best-known Cabibbo-Kobayashi-Maskawa (CKM) matrix element (as a
percentage of its value), if we assume three generations: $V_{tb} =
0.9990-0.9993$ \cite{Groom:2000in}. This is due to the small measured values
of $V_{ub}$ and $V_{cb}$ and the three-generation unitarity constraint
$|V_{ub}|^2 + |V_{cb}|^2 + |V_{tb}|^2 = 1$. Thus, if there are three
generations, we desire a measurement of $V_{tb}$ with an accuracy of $0.0002$.
Unfortunately, there is no known way to achieve such an accuracy.

If there are more than three generations, $V_{tb}$ is almost completely
unknown: $V_{tb} = 0.07-0.993$ \cite{Groom:2000in}.  In this case, a
measurement of $V_{tb}$ with any accuracy is worthwhile.  The existence of a
fourth generation is disfavored by precision electroweak data at the $97\%$
C.L., however \cite{Groom:2000in}.  If there are only three generations, then a
measurement of $V_{tb}$ may be considered as a probe of physics beyond the
standard model \cite{delAguila:2000aa,Tait:2001sh}.

CDF has measured \cite{Affolder:2000xb}
\begin{equation}
\frac{BR(t\to Wb)}{BR(t\to Wq)} =
\frac{|V_{tb}|^2}{|V_{td}|^2+|V_{ts}|^2+|V_{tb}|^2}=0.94^{+0.31}_{-0.24}
\label{ratio}
\end{equation}
and it is interesting to ask what this tells us about $V_{tb}$. If we assume
that there are just three generations of quarks, then unitarity of the CKM
matrix implies that the denominator of Eq.~(\ref{ratio}) is unity, and we can
immediately extract
\begin{equation}
|V_{tb}| = 0.97^{+0.16}_{-0.12} \;(> 0.75\;{\rm 95\%\;CL}) \;(3\;{\rm
generations}).
\end{equation}
However, we already know that $V_{tb} = 0.9990-0.9993$ if we assume three
generations, which is far more accurate than the above measurement. If we
assume more than three generations, then we lose the constraint that the
denominator of Eq.~(\ref{ratio}) is unity. All we can conclude from
Eq.~(\ref{ratio}) is that $|V_{tb}| >> |V_{ts}|,|V_{td}|$; we learn nothing
about its absolute magnitude.

\begin{figure}[t]
\begin{center}
\epsfxsize= 5.5in \leavevmode \epsfbox{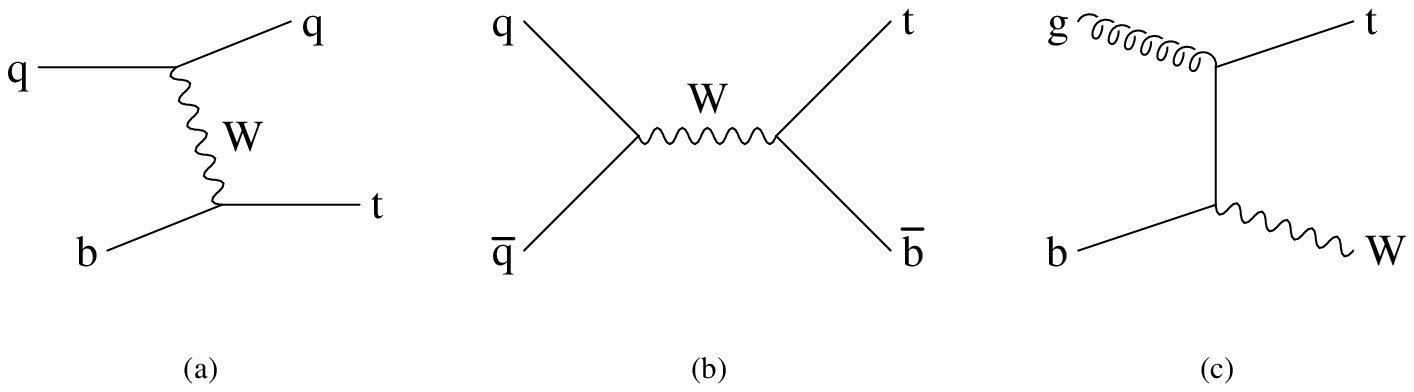}
\end{center}
\caption{Single-top-quark production via (a) $t$-channel $W$ exchange, (b)
$s$-channel $W$ exchange, and (c) associated production with a $W$.}
\label{singletop}
\end{figure}

Fortunately, there is a direct way to measure $|V_{tb}|$ at the Tevatron and
the LHC, which makes no assumptions about the number of generations.  One uses
the weak interaction to produce the top quark; the three relevant processes are
shown in Fig.~\ref{singletop}.  The cross sections for these ``single top''
processes are proportional to $|V_{tb}|^2$.  The first process involves a
$t$-channel $W$ boson \cite{Willenbrock:1986cr,Yuan:1990tc,Ellis:1992yw}, the
second process involves an $s$-channel $W$ boson
\cite{Cortese:1991fw,Stelzer:1995mi}, and the third process is associated
production of a single top quark with a real $W$ boson
\cite{Heinson:1997zm,Tait:2000cf}. The cross sections for these processes are
given in Table \ref{singletopsigma}, along with the cross section for $t\bar
t$ pair production. The largest single-top cross section comes from the
$t$-channel process, which is about 1/3 of the cross section for $t\bar t$
pair production. The $s$-channel process is relatively larger at the Tevatron
than the LHC, since it is a quark-antiquark initiated process. It has the
advantage of small theoretical uncertainty.  The cross section for associated
production is only significant at the LHC.

%%%%%%%%%%%%%%%%%%%%%%%%%%%%%%%%%%%%%%%%%%%%%%%%%%%%%%%%%%%%%%%%%%%%%%%%%
%%
%%   use this format to include a LaTeX table  into your paper
%%
\def\tableline{\noalign{%\vskip-.5pt
\hrule height.7pt depth0pt\vskip3pt}}

\begin{table}[htb]
\caption{Total cross sections (pb) for single-top-quark production and
top-quark pair production at the Tevatron and LHC, for $m_t = 175$ GeV. The
NLO $t$-channel cross section is from Ref.~\cite{Stelzer:1997ns}. The NNLO
$s$-channel cross section is from Refs.~\cite{Smith:1996ij,Chetyrkin:2000mq}.
The cross section for the $Wt$ process is from Ref.~\cite{Tait:2000cf}; it is
leading order, with a subset of the NLO corrections included.  The
uncertainties are due to variation of the factorization and renormalization
scales; uncertainty in the parton distribution functions; and uncertainty in
the top-quark mass (2 GeV).}\label{singletopsigma}
\begin{center}
\setlength{\tabcolsep}{9pt}
\renewcommand{\arraystretch}{1.2}
\begin{tabular}{lll}
\tableline & Tevatron & LHC \\ \tableline
$t$-channel  & $2.12 \pm 0.24$ & $238 \pm 27$ \\
$s$-channel & $0.88 \pm 0.06$ & $10.2 \pm 0.7$ \\
$Wt$ & $0.088 \pm 0.023$ & $51 \pm 9$ \\
$t\bar t$ & $\approx 6.5$ & $\approx 770$ \\
\hline
\end{tabular}
\end{center}
\end{table}

%%%%%%%%%%%%%%%%%%%%%%%%%%%%%%%%%%%%%%%%%%%%%%%%%%%%%%%%%%%%%%%%%%%%%%%%%

Thus far there are only upper bounds on the cross sections from CDF and D0.
The upper bounds on the $t$-channel cross sections are
\cite{Kikuchi:2000sv,Abbott:2001pa}
\begin{eqnarray}
\sigma(qb\to qt) & < & 13.5\;{\rm pb}\;({\rm 95\%\;CL})\;({\rm CDF})\\
\sigma(qb\to qt) & < & 58\;{\rm pb}\;({\rm 95\%\;CL})\;({\rm D0})
\end{eqnarray}
which is an order of magnitude away from the theoretical expectation. There is
a similar bound on the $s$-channel process \cite{Kikuchi:2000sv,Abbott:2001pa}
\begin{eqnarray}
\sigma(q\bar q \to t\bar b) &<& 12.9\;{\rm pb}\;({\rm 95\%\;CL})\;({\rm CDF})\\
\sigma(q\bar q \to t\bar b) &<& 39\;{\rm pb}\;({\rm 95\%\;CL})\;({\rm D0})
\end{eqnarray}
which is even further from the theoretical expectation. The $t$- and
$s$-channel processes will be first observed in Run II
\cite{Amidei:1996dt,Stelzer:1998ni,Belyaev:1999dn}, while the
associated-production process must await the LHC
\cite{ATLAS,Beneke:2000hk,Tait:2000cf}.

The most accurate measurements of $V_{tb}$ will come from the $t$- and
$s$-channel processes.  Both the $t$-channel \cite{Stelzer:1997ns} and the
$s$-channel \cite{Smith:1996ij} total cross sections have been calculated at
next-to-leading order (NLO) in QCD, with an uncertainty of about 5\% from
varying the factorization and renormalization scales.  After this symposium, a
calculation of the leading (in the large $N_c$ limit)
next-to-next-to-leading-order (NNLO) QCD correction to the $s$-channel process
appeared \cite{Chetyrkin:2000mq}; this essentially eliminates the uncertainty
from varying the factorization and renormalization scales. It is also
desirable to have a calculation of the differential cross section at NLO; this
work is in progress \cite{Harris:2000sv}. Taking all uncertainties into
account, it seems possible that $V_{tb}$ can be measured at the Tevatron and
the LHC with an uncertainty of 5\% \cite{Beneke:2000hk,Stelzer:1998ni}.

At a Linear Collider, $V_{tb}$ can be extracted by measuring the top-quark
width from a scan of the $t\bar t$ threshold.  The anticipated uncertainty in
$V_{tb}$ from such a measurement is about $10\%$ \cite{Comas:1996kt}. The
width is known with very good theoretical precision, thanks to recent
calculations at NNLO in QCD \cite{Czarnecki:1999qc,Chetyrkin:1999ju}. The
recent renormalization-group-improved calculation of the $t\bar t$ threshold
(Fig.~\ref{thresh})), mentioned in the previous section, removes any
theoretical uncertainty in the normalization of the cross section that would
impede the extraction of the width \cite{Hoang:2000ib}.

\section{Yukawa coupling}

As mentioned in the introduction, the top-quark Yukawa coupling is related to
the top-quark mass (at leading order) via $y_t=\sqrt 2 m_t/v$, where $v$ is
the vacuum-expectation value of the Higgs field.  However, it is the Yukawa
coupling, not the mass, which is the truly fundamental parameter.  The Yukawa
coupling transmits the information that the Higgs field has acquired a vacuum
expectation value to the top quark, thereby generating its mass.  Since the
Yukawa coupling is associated both with electroweak symmetry breaking and with
fermion mass generation, it may be the most interesting parameter in top-quark
physics.

How accurately do we desire to measure the top-quark Yukawa coupling?  Since it
is linearly related to the top quark mass, it would be desirable to measure it
with the same fractional precision as the top-quark mass.  A measurement of
the top-quark mass with an accuracy of 1--2 GeV would correspond to a
measurement of the Yukawa coupling to about 1\%.  Unfortunately, there is no
known way to make a measurement with this accuracy.

\begin{figure}[t]
\begin{center}
\epsfxsize= 2.5in \leavevmode \epsfbox{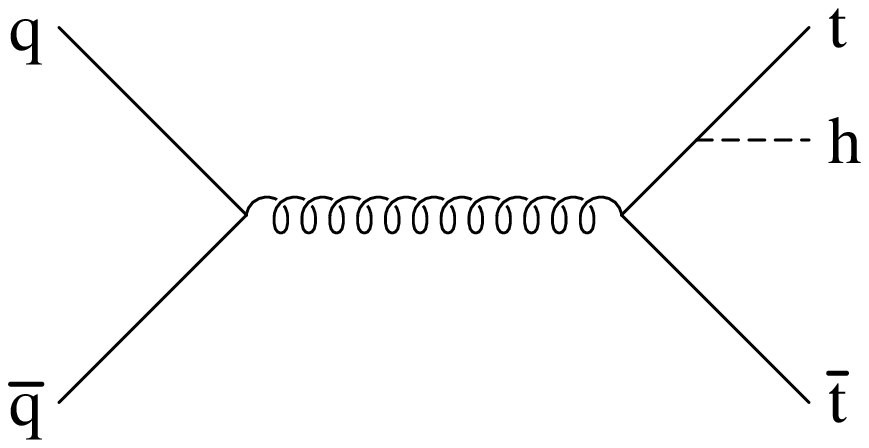}
\end{center}
\caption{Higgs-boson production in association with a top-quark pair.}
\label{tth}
\end{figure}

The most direct way to measure the top-quark Yukawa coupling at a hadron
collider is via the associated production of the Higgs boson with a top-quark
pair, as shown in Fig.~\ref{tth}.  If the Higgs boson decays to $b\bar b$,
there are four $b$ quarks in the final state, and tagging three or more of
them reduces the background to an acceptable level.  It has recently been
argued that this process can be used to discover the Higgs boson in Run II of
the Tevatron, given 15 fb$^{-1}$ of integrated luminosity
\cite{Goldstein:2000bp}.  This process would yield only a crude measurement of
the Yukawa coupling, however, due to the limited statistics. Even at the LHC,
the anticipated accuracy is only about 16\% via this process
\cite{ATLAS,Beneke:2000hk}.  A next-to-leading-order calculation of the
production cross section is still needed. This calculation has thus far been
performed only in the limit $m_h \ll m_t$ \cite{Dawson:1998im}.

\begin{figure}[t]
\begin{center}
\epsfxsize= 2.0in \leavevmode \epsfbox{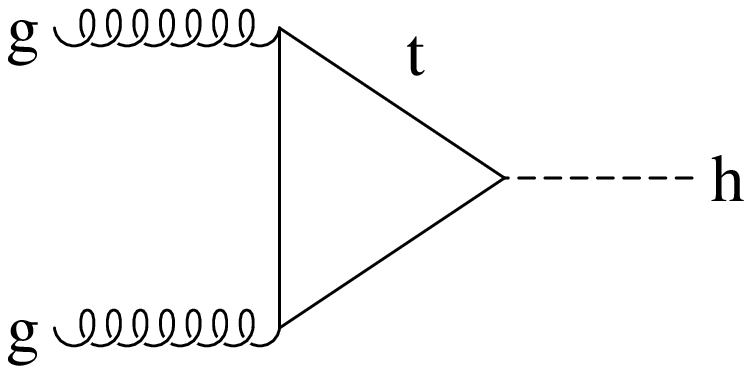}
\end{center}
\caption{Higgs-boson production from gluon fusion via a top-quark loop.}
\label{ggh}
\end{figure}

A less direct way to measure the top-quark Yukawa coupling at a hadron
collider is to produce the Higgs boson via gluon fusion, as shown in
Fig.~\ref{ggh} \cite{Georgi:1978gs}.  In the standard model this process is
dominated by a top-quark loop, but if there are other heavy colored particles
that couple to the Higgs boson (such as squarks), they too contribute to the
amplitude, complicating the extraction of the top-quark Yukawa coupling.
Assuming only the top quark contributes substantially to this process, the
Yukawa coupling can be measured with an accuracy of about 10\%.  The
next-to-leading order calculation of this cross section is already in hand,
but the remaining scale dependence is significant, about $15\%$
\cite{Spira:1995rr}.  The gluon luminosity contributes another $10\%$ to the
uncertainty in the cross section \cite{Huston:1998jj}.

The top-quark Yukawa coupling can also be measured at a Linear Collider, using
the analogue of Fig.~\ref{tth} with the initial quark-antiquark replaced by
electron-positron and the intermediate gluon replaced by $\gamma,Z$.  The
measurement is limited by statistics, and depends on the machine energy.  In
Table \ref{tthlc} I list the accuracy of the measurement of the Yukawa
coupling at a 500 GeV and a 1 TeV Linear Collider for two Higgs masses
\cite{Baer:2000ge}.  At the 500 GeV machine one is limited by phase space. One
does much better at the 1 TeV machine, but cannot achieve the desired 1\%
accuracy. The next-to-leading-order calculation of this cross section is
already available \cite{Dawson:1999ej,Dittmaier:1998dz}.

%%%%%%%%%%%%%%%%%%%%%%%%%%%%%%%%%%%%%%%%%%%%%%%%%%%%%%%%%%%%%%%%%%%%%%%%%
%%
%%   use this format to include a LaTeX table  into your paper
%%
\def\tableline{\noalign{%\vskip-.5pt
\hrule height.7pt depth0pt\vskip3pt}}

\begin{table}[htb]
\caption{Accuracy of the measurement of the top-quark Yukawa coupling from
$e^+e^- \to t\bar th$ at a Linear Collider of energy 500 GeV and 1 TeV. From
Ref.~\cite{Baer:2000ge}.}\label{tthlc}
\begin{center}
\setlength{\tabcolsep}{9pt}
\renewcommand{\arraystretch}{1.2}
\begin{tabular}{lll}
\tableline $m_h$ & 500 GeV & 1 TeV \\ \tableline
110 GeV & 12\% & 6\% \\
130 GeV & 44\% & 8\% \\
\hline
\end{tabular}
\end{center}
\end{table}

%%%%%%%%%%%%%%%%%%%%%%%%%%%%%%%%%%%%%%%%%%%%%%%%%%%%%%%%%%%%%%%%%%%%%%%%%

\newpage

\section{Conclusions}

Thus far the properties of the top quark have been tested only crudely.  This
decade will witness measurements of the top quark with increasing precision at
the Tevatron and the LHC, and perhaps eventually at a Linear Collider.  These
measurements will either confirm that the top quark is an ordinary
standard-model quark, or will indicate the presence of new physics.  In either
case, the study of the top quark will be rewarding.

In this talk I concentrated on the measurement of the fundamental parameters
associated with the top quark.  I argued that the desire to measure the
top-quark mass to an accuracy of 1--3 GeV, the goal of the Tevatron and LHC, is
well motivated by precision electroweak analyses, and by comparison with the
anticipated accuracy in the $b$-quark mass.  Although the Linear Collider can
measure the mass with an accuracy of 100 MeV, there is no compelling
motivation within the standard model to pursue such an accuracy.  I considered
two speculative motivations for pursuing this accuracy from physics beyond the
standard model.

The CKM matrix element $V_{tb}$ will be measured with an accuracy of about
$5\%$ at the Tevatron and the LHC via single-top-quark production.  If there
are only three generations, this measurement is not nearly accurate enough to
help determine the CKM matrix.  A Linear Collider cannot make a more accurate
measurement.

The top-quark Yukawa coupling to the Higgs boson is perhaps the most
interesting parameter, since it is associated with electroweak symmetry
breaking and fermion mass generation.  Since the Yukawa coupling is
proportional to the top-quark mass, it would be desirable to measure them both
with the same percentage accuracy.  Unfortunately, this is well out of reach.
Only a crude measurement of the top-quark Yukawa coupling can be made at the
LHC.  A Linear Collider with energy significantly above 500 GeV can measure
the Yukawa coupling with an accuracy below $10\%$, but cannot achieve the
desired $1\%$ accuracy.

\Acknowledgments I am grateful for conversations with and assistance from
S.~Heinemeyer, A.~Hoang, and K.~Paul. This work was supported in part by
Department of Energy grant DE-FG02-91ER40677.

\end{document}